\documentclass[conference]{IEEEtran}
\usepackage{threeparttable}
\usepackage[caption=false]{subfig}
\usepackage{booktabs} % 添加这一行
\usepackage{cite}
\usepackage{amsmath,amssymb,amsfonts}
\usepackage{algorithmic}
\usepackage{graphicx}
\usepackage{textcomp}
\usepackage{xcolor}
\def\BibTeX{{\rm B\kern-.05em{\sc i\kern-.025em b}\kern-.08em
    T\kern-.1667em\lower.7ex\hbox{E}\kern-.125emX}}
\begin{document}

\title{A Novel Grant Prediction Method for 5G NR Terminals\\
% {\footnotesize \textsuperscript{*}Note: Sub-titles are not captured for https://ieeexplore.ieee.org  and
% should not be used}
% \thanks{Identify applicable funding agency here. If none, delete this.}
}

\author{\IEEEauthorblockN{Chenhao Wu, Xiaojiang Xu, Yuxuan Li, Yuanhao Xu, Wenhui Xiong, and Xiaoyu Fu}
\IEEEauthorblockA{National Key Laboratory of Wireless Communications,\\ University of Electronic Science and Technology of China, Chengdu, China  611731}
% Emails: 202322220521@std.uestc.edu.cn,202222220521@std.uestc.edu.cn??, 202321220520@std.uestc.edu.cn,202421220518@std.uestc.edu.cn, whxiong@uestc.edu.cn, ffu@uestc.edu.cn \quad (*Corresponding author)
}
\maketitle

\begin{abstract}
5G NR user equipment suffers from high power consumption due to 
continuous PDCCH monitoring. Predictive dynamic power management (DPM) 
can save energy by forecasting data grants, but accurate 
prediction is challenging due to unobservable 
scheduling states and bursty grant patterns. 
This paper proposes IOHMM-BO, a high-order input-output hidden 
Markov model with Bayesian optimization. Based on real 5G NR traces, 
we capture long-range dependencies via a compound state and jointly 
optimize model order and listening window using Bayesian 
optimization. Experiments on real traces show that IOHMM-BO 
achieves 45.3\% accuracy, 5.0\% false negative rate, 
and 43\% energy saving with low 
computational overhead. The method 
provides a balanced trade-off between reliability and 
energy efficiency.
\end{abstract}

\begin{IEEEkeywords}
5G NR, grant prediction, dynamic power management, input-output hidden Markov model, Bayesian optimization.
\end{IEEEkeywords}

\section{Introduction}

With the widespread deployment of 5G systems, the power consumption of user equipment (UE) has become a critical concern. In cellular communications, a UE must continuously perform blind decoding on the physical downlink control channel (PDCCH) to obtain downlink control information (DCI), even when no data grant is scheduled. This mandatory monitoring process results in substantial energy waste \cite{wang2014end}. As illustrated in \cite{mo2024network,lin2023survey}, the fraction of time without valid data grants can exceed 90\% in typical mobile applications, yet the UE cannot foresee the arrival of grants and therefore keeps its receiver chain active unnecessarily.

Dynamic power management (DPM) techniques offer a promising solution. Predictive (or proactive) DPM aims to forecast whether a grant will appear in future transmission time intervals (TTIs) and switches the receiver to a low-power mode when no grant is predicted, thereby reducing energy consumption \cite{brand2021multi,lin2023survey}.
However, realizing effective predictive DPM in 5G NR is challenging for two main reasons: the scheduling state at the gNB side is only partially observable to the UE, and the bursty nature of grants makes them difficult to predict accurately.

\subsection{Related Work}

Research on UE energy saving can be broadly classified into two categories. 
The first category focuses on base station (gNB) cooperation, e.g., by optimizing discontinuous reception 
(DRX) parameters \cite{A_Survey_on_DRX_Mechanism,Power_Saving_Techniques,Accurate_Modeling_of_the_DRX} or 
introducing wake-up signals \cite{Wake-up_Radio-Based_5G_Mobile_Access}. Some works employ traffic prediction 
to adjust DRX cycles adaptively \cite{zhou2019online,azari2022energy, wang2024base,Traffic_Driven_Sounding,Energy_Management_in}. 
While these methods are protocol-compatible, they often require additional signaling or rely on 
coarse-grained UE state knowledge.

The second category deploys prediction algorithms directly on the UE side, to which this paper belongs.
Early attempts used traditional machine learning models such as multilayer perceptrons,
 random forests, and k-nearest neighbors to predict LTE/NR grants 
 \cite{brand2021adaptive,sue2016binary,brand2020clustering,brand2021multi}. More recent work has explored deep 
 learning \cite{mo2024network} and reinforcement learning \cite{lin2023survey}. 
 These studies mainly focus on applying existing machine learning methods to grant prediction.In contrast, this paper makes specific improvements based on the characteristics of real‑network grant data, thereby achieving better performance.

\subsection{Contributions}

\begin{figure}[htbp]
    \centering
    \subfloat[]{
        \includegraphics[width=0.41\textwidth]{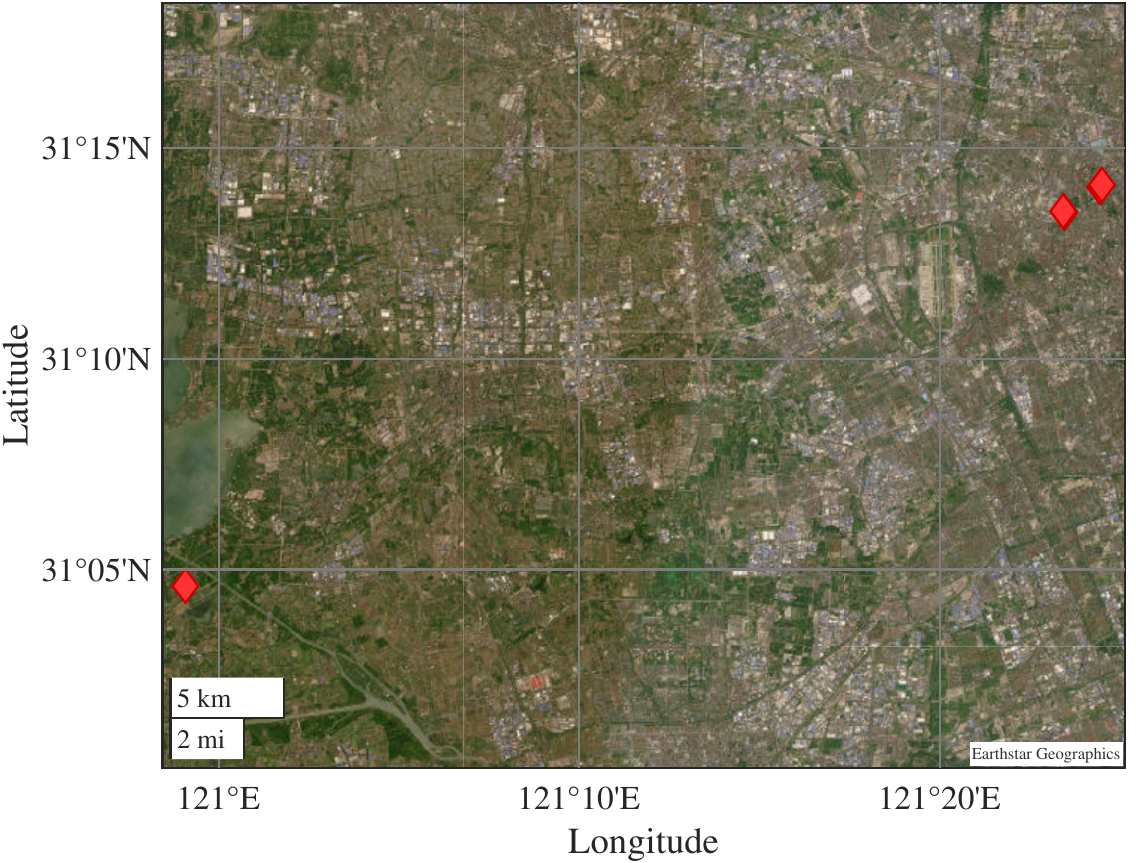}
    }
    \hfill
    \subfloat[]{
        \includegraphics[width=0.41\textwidth]{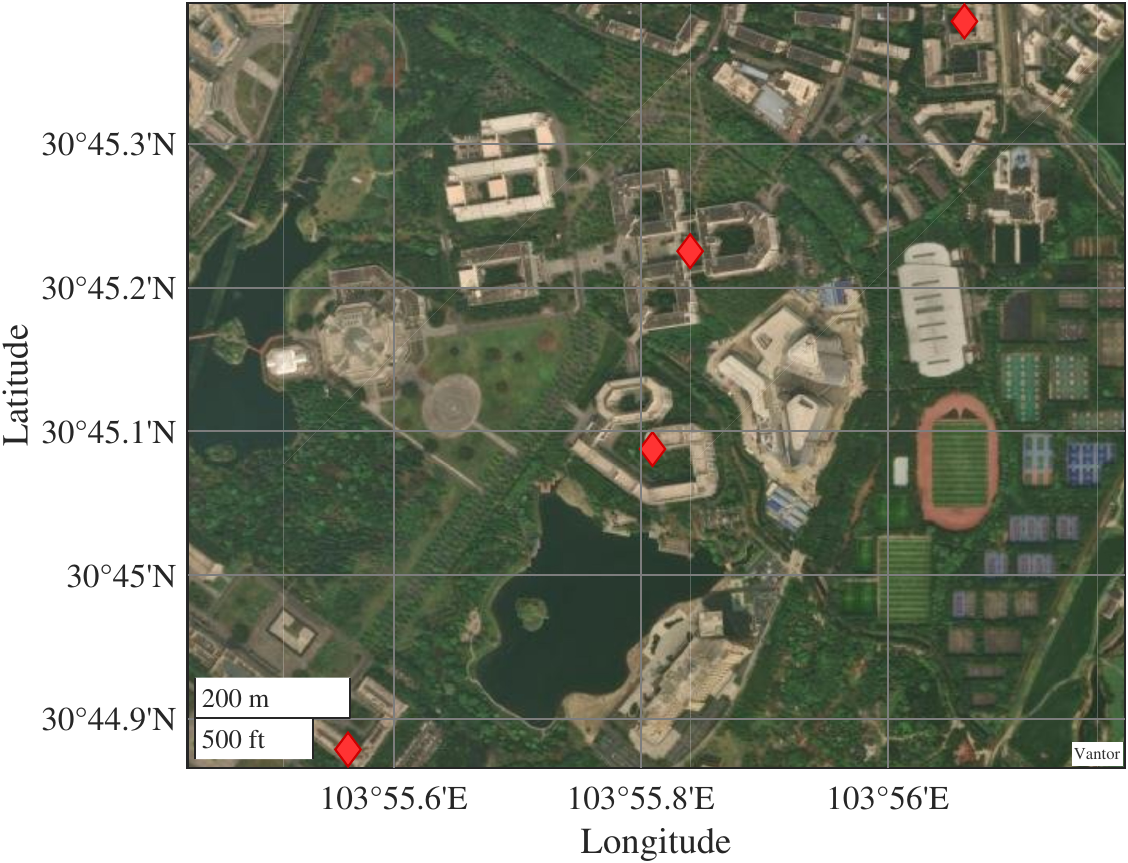}
    }
    \caption{Geographic distribution of data collection locations in two cities.
    Red markers indicate the positions where 5G NR traces were collected.}
    \label{fig:geo_distribution}
\end{figure}

In this paper, we propose a novel predictive DPM method based on a high-order input-output hidden Markov model (IO-HMM) with Bayesian optimization (BO), referred to as IOHMM-BO. The main contributions are:

\begin{itemize}
    \item \textbf{Data-driven analysis:} We collect real-world 5G NR traces covering six popular applications
    (TikTok, Honor of Kings, Amap, WeChat Short Video, iReader, and NetEase Cloud Music) and several representative scenarios 
    (shopping mall, office, outdoor). 
    Field measurements are conducted at multiple sites across two cities with 
    distinct spatial characteristics; as illustrated in Fig.~\ref{fig:geo_distribution}, red markers indicate the locations of our sampling sites. 
    This dataset ensures sufficient spatial diversity for evaluating the performance of predictive DPM schemes.

    \item \textbf{High-order IO-HMM with joint optimization:} To capture the unobservable scheduling state and long memory of grant sequences, we formulate the grant generation process as a high-order IO-HMM and transform it into an equivalent first-order form via compound state aggregation. The model order \(K\), monitoring window \(Wdz\), and IO-HMM parameters are jointly optimized using a two-level framework (inner EM for parameter estimation and outer Bayesian optimization with an accuracy-FNR trade-off), automatically balancing prediction accuracy and miss-detection risk.
    
    \item \textbf{Realistic energy evaluation:} We adopt a receiver power model based on power state machines and quantify the inherent computation cost of the algorithm via FLOP-to-energy mapping. Normalized energy consumption is used to compare predictive DPM against reactive DPM.
\end{itemize}

Experimental results on real 5G NR datasets show that IOHMM-BO significantly outperforms baseline methods (FNN, DQN, etc.) in terms of accuracy, false negative rate (FNR) control, and net energy saving, while maintaining low computational overhead.

\subsection{Paper Organization}

The remainder of this paper is structured as follows. Section II describes the system model and the proposed IOHMM-BO method, including the high-order IOHMM formulation, equivalent transformation, EM training, and Bayesian optimization of hyperparameters. Section III presents the experimental setup, evaluation metrics (accuracy, FNR, normalized energy), and results. Finally, Section IV concludes the paper and outlines future directions.

\section{Data Analysis and Problem Formulation}
\subsection{Statistical Characteristics of Grant Sequences}

Define the grant indicator at TTI \(t\) as
\[
G_t \in \{0,1\},
\]
where \(G_t=1\) means a valid downlink grant is detected. Our analysis reveals two key characteristics.

\subsubsection{Bursty Clusters}

\begin{figure}[htbp]
\centering
\includegraphics[width=0.45\textwidth]{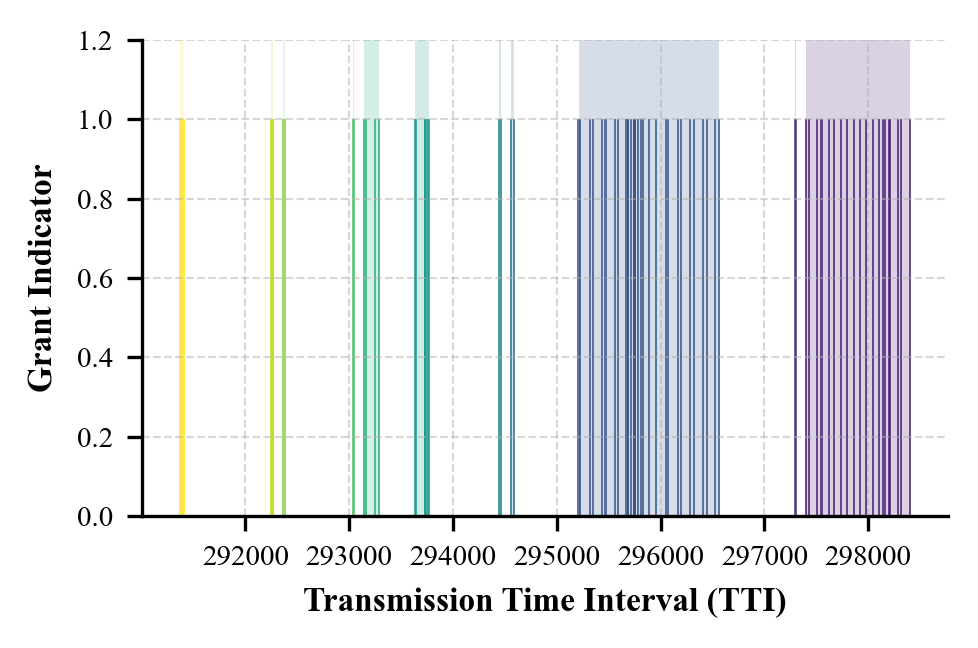}
\caption{Bursty clusters of grants in a representative trace.}
\label{fig:clusters}
\end{figure}

As shown in Fig.~\ref{fig:clusters}, grants do not appear uniformly but 
in bursty clusters: several consecutive or densely packed grants followed 
by long silent periods. Each cluster typically corresponds to an active 
data transmission phase. This phenomenon arises because the gNB 
scheduler, upon detecting queued data, tends to allocate resources 
continuously for a period.
This observation implies that the presence of a grant at 
time \(t\) strongly depends on past grants, i.e., the process has 
high-order temporal memory.

\subsubsection{Long-Range Influence of Scheduling Requests (SR)}

\begin{figure}[htbp]
\centering
\includegraphics[width=0.44\textwidth]{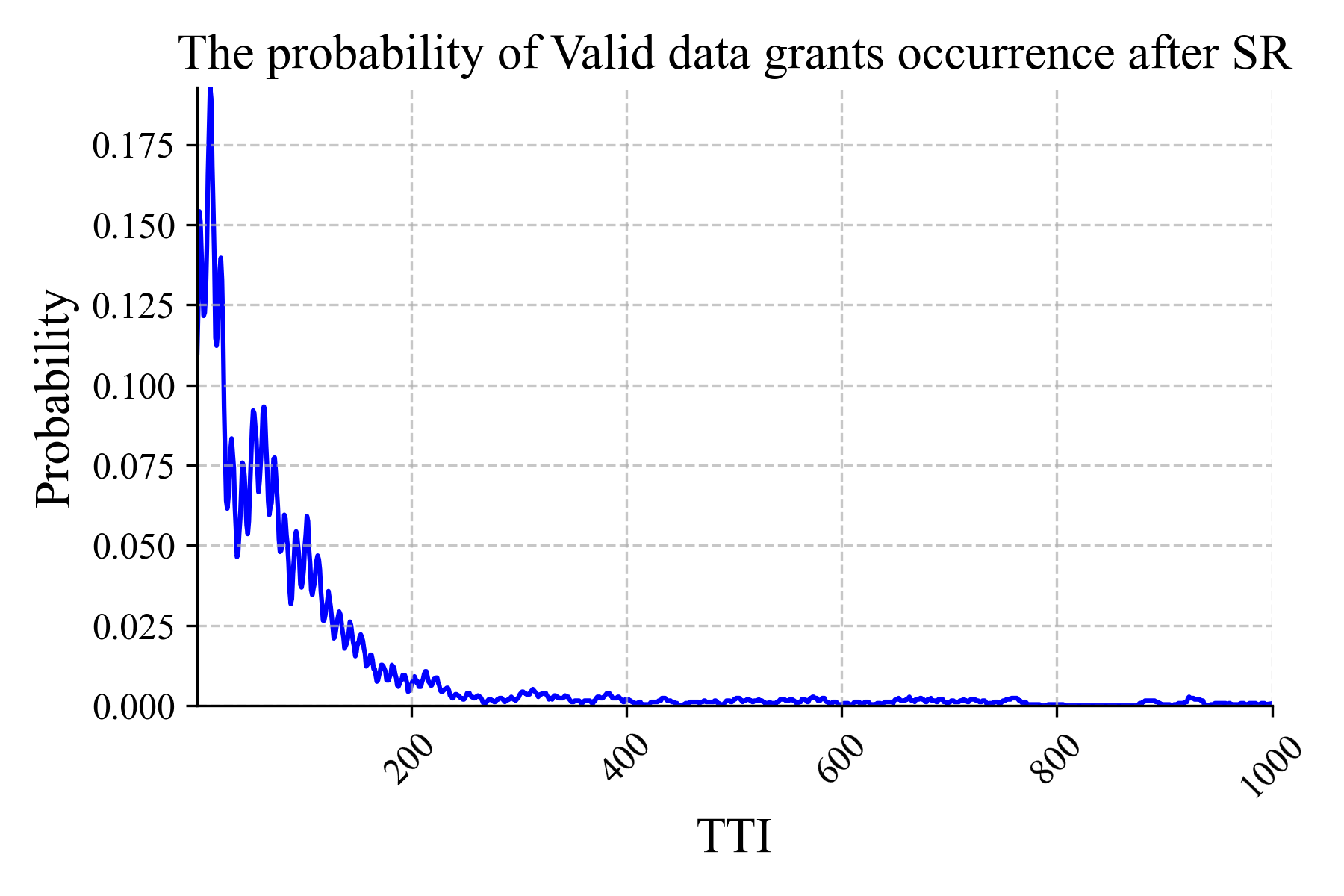}
\caption{Conditional probability of a downlink grant after an SR event.}
\label{fig:sr_grant_probability}
\end{figure}

A scheduling request (SR) is sent by the UE when new uplink data arrives. 
Fig.~\ref{fig:sr_grant_probability} plots the conditional 
probability \(P(G_{t+\Delta t}=1 \mid SR_t)\). The probability 
rises sharply immediately after an SR and then decays gradually 
over tens of TTIs.
Thus, SR events are strong indicators of future 
grants and their influence persists over multiple TTIs.

Together, these two observations indicate that the grant arrival process 
is neither memoryless nor independent. The bursty cluster pattern calls 
for a model that retains information across multiple TTIs, while the 
prolonged effect of SR events suggests that external signaling should 
be explicitly incorporated as input features. These requirements 
motivate the use of a high-order input-output hidden Markov model 
(IO-HMM).

\subsection{Problem Formulation}
The gNB scheduler's internal state (buffer status, channel quality, other UEs' competition) is not observable to the UE. However, the UE can observe local features: \(SR_t\), time distance to the last SR, CDRX state transitions, and time to last CDRX rising edge. We assemble these into an input vector
\begin{equation}
\begin{aligned}
U_t = \big[ & SR_t,\; \exp(-0.01\cdot\Delta SR_t), \\
& Cdrx_t^{\text{rise}},\; \exp(-0.01\cdot\Delta Cdrx^{\text{rise}}) \big].
\end{aligned}
\end{equation}

We abstract the unobservable scheduler state by a binary hidden variable \(\tilde S_t \in \{0,1\}\), representing a low-activity (0) or high-activity (1) scheduling phase. The observable grant indicator at each TTI is defined as \(G_t \in \{0,1\}\), where \(G_t = 1\) indicates a valid downlink grant and \(G_t = 0\) otherwise. The sequence \(\{G_t\}\) serves as the observation sequence. Given the bursty clusters and long-range SR influence, the grant generation process exhibits high-order dependencies. Therefore, we model it with a high-order input-output hidden Markov model (IO-HMM).

\section{Proposed IOHMM-BO Method}

\subsection{High-Order IO-HMM and Equivalent First-Order Form}
Let the model order be \(K\). The high-order state transition and observation dependencies are
\begin{align}
\begin{split}
&P(\tilde S_t \mid \tilde S_{1:t-1}, G_{1:t-1}, U_{1:t}) \\
&= P(\tilde S_t \mid \tilde S_{t-1}, \dots, \tilde S_{t-K},\; \tilde U_t^{(K)}),
\end{split} \\
\begin{split}
&P(G_t \mid \tilde S_{1:t}, G_{1:t-1}, U_{1:t}) \\
&= P(G_t \mid \tilde S_t, \dots, \tilde S_{t-K+1},\; \tilde U_t^{(K)}),
\end{split}
\end{align}
where \(\tilde U_t^{(K)} = [U_t, U_{t-1}, \dots, U_{t-K}]\) is the stacked input over the last \(K\) TTIs.

To enable standard inference, following the method of \cite{High_order_HMM}, we define a compound state
\[
Q_t = [\tilde S_t, \tilde S_{t-1}, \dots, \tilde S_{t-K+1}]^\top \in \mathcal{Q}_K,\quad |\mathcal{Q}_K| = 2^K.
\]
Then the model becomes a first-order IO-HMM:
\[
P(Q_t \mid Q_{t-1}, \tilde U_t^{(K)}),\qquad P(G_t \mid Q_t, \tilde U_t^{(K)}).
\]

\subsection{Parameterization and EM Training}

The input feature map is \(\phi(\tilde U_t^{(K)}) = [1,\; \mathrm{Norm}(\tilde U_t^{(K)})]\). For each compound state \(j\), the emission probability is Bernoulli:
\[
P(G_t=1 \mid Q_t=j, \tilde U_t^{(K)}) = \sigma\big(\boldsymbol{\beta}_j^\top \phi(\tilde U_t^{(K)})\big),
\]
with \(\sigma(x)=1/(1+e^{-x})\). For transition, given current state \(i\), only two successor states are valid (corresponding to \(\tilde S_t=1\) or \(0\)), denoted \(q_i^{(1)}\) and \(q_i^{(0)}\). The transition probability is
\[
P(Q_t = q_i^{(1)} \mid Q_{t-1}=i, \tilde U_t^{(K)}) = \sigma\big(\boldsymbol{w}_i^\top \phi(\tilde U_t^{(K)})\big).
\]

The model parameters \(\Theta_K = \{\Pi, \{\boldsymbol{w}_i\}, \{\boldsymbol{\beta}_j\}\}\) are estimated via the EM algorithm. In the E-step, the forward–backward algorithm computes posterior state probabilities. In the M-step, emission and transition parameters are updated by solving weighted logistic regression problems.

\subsection{Joint Hyperparameter Optimization via Bayesian Optimization}

The model order \(K\) and the maximum listening window length \(Wdz\) (a hyperparameter controlling the upper bound of consecutive monitoring TTIs) are jointly optimized by a two-level framework:
\begin{itemize}
    \item \textbf{Inner level}: For fixed \((K, Wdz)\), estimate \(\Theta_K\) on the training set using EM.
    \item \textbf{Outer level}: Evaluate performance on a validation set using 
    \(\mathcal{L}_{\mathrm{outer}} = -\mathrm{ACC} + \beta \cdot \max(0, \mathrm{FNR} - thr)\), 
    where ACC is accuracy, FNR is false negative rate, \(thr\) is a tolerated FNR threshold, and \(\beta\) a penalty coefficient.
\end{itemize}
We adopt Bayesian optimization (BO) with a Gaussian process surrogate and expected improvement (EI) acquisition function \cite{Efficient_Global_Optimization,Snoek2012BO}.

\begin{figure}[htbp]
    \centering
    \subfloat[]{
        \includegraphics[width=0.36\textwidth]{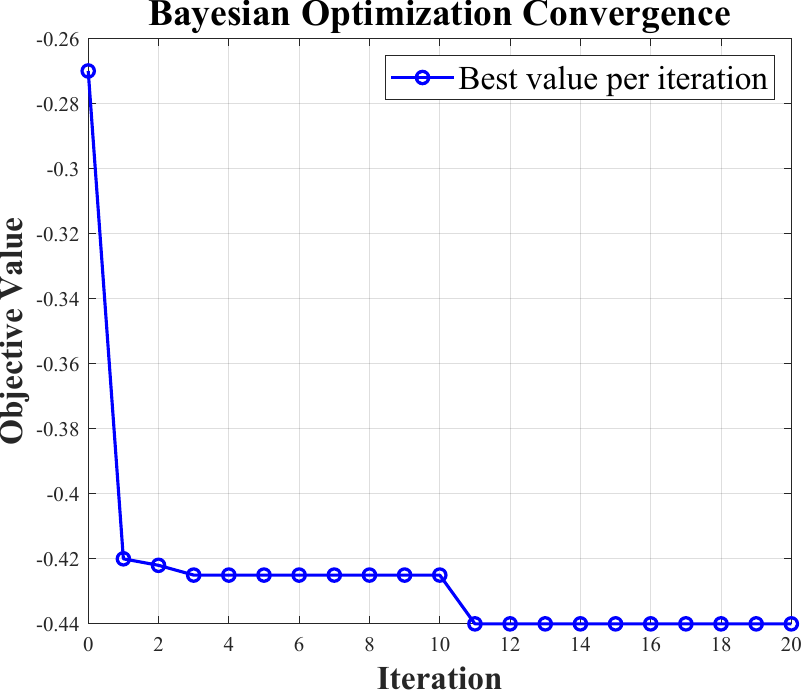}
        \label{fig:bayes1}
    }
    \hfill
    \subfloat[]{
        \includegraphics[width=0.38\textwidth]{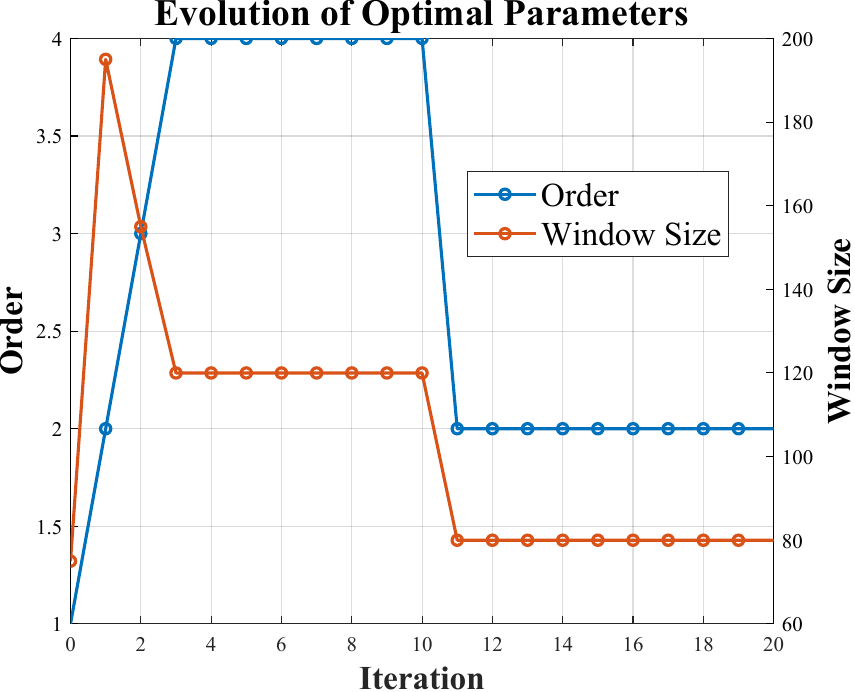}
        \label{fig:bayes2}
    }
    \caption{Bayesian optimization process on a TikTok trace: (a) objective value convergence; (b) hyperparameter trajectories.}
    \label{fig:bayes}
\end{figure}

Fig.~\ref{fig:bayes} illustrates the BO convergence on a representative trace (TikTok). 
Subfigure~\ref{fig:bayes1} shows the evolution of the objective 
value \(\mathcal{L}_{\mathrm{outer}}\) over BO iterations, which decreases and stabilizes within 20 iterations. 
Subfigure~\ref{fig:bayes2} displays the corresponding variations of the two hyperparameters \((K, Wdz)\), 
showing that the optimizer gradually settles on \(K=2\) and \(Wdz=78\) after an initial exploration phase.

\subsection{Online Prediction}

After training, we compute the online state distribution \(\alpha_t(j) = P(Q_t=j \mid \tilde U_{1:t}^{(K)})\) recursively:
\[
\alpha_t(j) = \sum_{i\in\mathcal{Q}_K} \alpha_{t-1}(i)\, a_{ij}(t),
\]
with initialization \(\alpha_0(j)=\pi_j\). The grant prediction probability is
\[
\hat p_t = \sum_{j\in\mathcal{Q}_K} \alpha_t(j)\, b_j^{(1)}(t).
\]

Using a fixed threshold may cause frequent fluctuations and increase missed grants. 
Therefore, we adopt a probability-driven continuous listening policy: a larger \(\hat p_t\) implies
 a more active scheduling phase, requiring a longer listening window to avoid missing grants;
a smaller \(\hat p_t\) suggests a sparse phase, where a shorter window suffices to save energy. 
Let \(Wdz\) be the maximum window length. The adaptive listening duration is \(d_t = \hat p_t \times Wdz\). 
This continuous adjustment balances miss-detection risk and energy consumption. The local CDRX state 
is enforced as a hard constraint: no monitoring during CDRX sleep.

\subsection{Energy Consumption Model}
\label{subsec:Energymodel}

We adopt a receiver power model based on power state machines \cite{wang2014end,A_Fine_grained}. For each TTI \(n\) and component \(m \in \{\text{RF\_RX}, \text{PHY\_RX}\}\), let \(t_{m,\text{on}}[n]\) and \(t_{m,\text{off}}[n]\) be the durations for which the component is in high-power (on) and low-power (off) states, respectively, with corresponding power consumptions \(P_{\text{on}}^m\) and \(P_{\text{off}}^m\). The total energy for predictive DPM over \(N\) TTIs is
\begin{equation}
\begin{aligned}
&E_{p\_\text{DPM\_RX}} = \sum_{n=1}^{N} \sum_{m \in \{\text{RF\_RX},\text{PHY\_RX}\}} \big( t_{m,\text{on}}[n] P_{\text{on}}^m \\
&+ t_{m,\text{off}}[n] P_{\text{off}}^m \big) + E_Q,
\end{aligned}
\end{equation}
where \(E_Q\) is the algorithm's inherent computation energy. To quantify \(E_Q\), we assume the algorithm runs on a baseband digital signal processor (DSP) with clock frequency \(f = 300\ \text{MHz}\) and power consumption \(P_{\text{DSP}} = 300\ \text{mW}\). Assuming one FLOP per clock cycle and a TTI duration \(t_{\text{slot}} = 0.5\ \text{ms}\), the maximum FLOPs per TTI is
\begin{equation}
\begin{aligned}
C_{\text{max}} &= f \times t_{\text{slot}} \\
&= 300 \times 10^6 \times 0.5 \times 10^{-3} = 150,000\ \text{FLOP}.
\end{aligned}
\end{equation}
Let \(C_Q\) be the algorithm's FLOP count per TTI (e.g., from Table~\ref{tab:perf}). The computation energy per TTI is
\begin{equation}
E_Q[n] = \frac{C_Q}{C_{\text{max}}} \cdot P_{\text{DSP}} \cdot t_{\text{slot}} = C_Q \cdot 1.0 \times 10^{-6}\ \text{mJ},
\end{equation}
and the total inherent energy is \(E_Q = \sum_{n=1}^{N} E_Q[n]\).

Normalized energy is defined as \(N_{\text{energy}} = E_{p\_\text{DPM\_RX}} / E_{r\_\text{DPM\_RX}}\), where \(E_{r\_\text{DPM\_RX}}\) is the energy of reactive DPM. This model enables fair comparison of different prediction algorithms by accounting for both receiver operation and computation overhead.

\section{Evaluation}

\subsection{Experimental Setup}

\subsubsection{Dataset}

We collected real 5G NR traces from commercial networks covering six popular applications (TikTok, Honor of Kings, WeChat Video, NetEase Music, iReader, and Gaode Map) and four typical scenarios (indoor mall, office, outdoor mall, and mobile environment). The dataset comprises 80 independent sessions, each about 5 minutes long. Table~\ref{tab:dataset_summary} summarizes the distribution. The grant indicator \(G_t\) is highly imbalanced: the fraction of TTIs with a valid downlink grant is around 10\% on average.

\begin{table}[htbp]
\centering
\caption{Dataset distribution by application and scenario}
\label{tab:dataset_summary}
\footnotesize
\begin{tabular}{lcccc}
\toprule
Application & Indoor mall & Office & Outdoor mall & Mobile \\
\midrule
TikTok          & 2 & 7 & 4 & 2 \\
Gaode Map    & 4 & 9 & 4 & 0 \\
Honor of Kings  & 4 & 10 & 4 & 1 \\
NetEase Music   & 4 & 1 & 2 & 2 \\
WeChat Video    & 4 & 2 & 2 & 1 \\
iReader        & 6 & 1 & 3 & 1 \\
\midrule
Total           & 24 & 30 & 19 & 7 \\
\bottomrule
\end{tabular}
\end{table}

\subsubsection{Baseline Methods and Evaluation Metrics}

We compare the proposed IOHMM-BO against three baselines:
\begin{itemize}
    \item \textbf{FFNN}: A three-layer feedforward neural network with input window of 10 TTIs and 6 features per TTI, trained with a cost-sensitive loss to handle class imbalance \cite{brand2021adaptive}.
    \item \textbf{DQN}: A deep Q-network with two-layer MLP and experience replay, using a 15-TTI sliding window of 6 features as state \cite{brand2021adaptive}.
    \item \textbf{SARSA-$\lambda$}: A tabular reinforcement learning method with eligibility traces, using the last 10 TTIs of grant observations as state \cite{Reinforcement_Learning_for}.
\end{itemize}
The reactive DPM
%(legacy scheme)
 is used as the energy baseline.

The following metrics are used:
\begin{itemize}
    \item \textbf{Accuracy (ACC)}: Overall fraction of correctly predicted TTIs.
    \item \textbf{False Negative Rate (FNR)}: Fraction of grants that are missed. This is the primary reliability constraint.
    \item \textbf{Normalized Energy (\(N_{\text{energy}}\))}: Defined in Section \ref{subsec:Energymodel}, comparing predictive DPM against reactive DPM including algorithm computation cost.
    \item \textbf{Computational overhead}: Estimated via FLOP count per TTI.
\end{itemize}

% All algorithms are trained on 70\% of the sessions and tested on the remaining 30\%. For IOHMM-BO, the hyperparameter search spaces are \(K \in \{2,3,4,5,6\}\) and \(Wdz \in \{1,2,\dots,10\}\). The outer objective uses \(thr=0.08\) and \(\beta=100\).
All algorithms are trained on 70\% of the sessions and tested on the remaining 30\%. For IOHMM-BO, the hyperparameter search spaces are \(K \in \{2,3,4,5,6\}\), while the search range for \(Wdz\) is adaptively determined according to each application. The outer objective uses \(thr=0.08\) and \(\beta=100\).

\subsection{Prediction Performance}

Table~\ref{tab:perf} reports the average ACC and FNR over all test scenarios. IOHMM-BO achieves the highest average accuracy (45.3\%) and a low FNR (5.0\%), outperforming FFNN (21.7\% ACC, 4.6\% FNR), DQN (40.7\% ACC, 23.2\% FNR), and SARSA-$\lambda$ (31.1\% ACC, 24.7\% FNR). Notably, while FFNN has a slightly lower FNR (4.6\%) than IOHMM-BO (5.0\%), its accuracy is much lower, indicating that it predicts “no grant” almost always, failing to capture actual grants. IOHMM-BO strikes a better balance between detecting grants and avoiding false alarms.

\begin{table}[htbp]
\centering
\caption{Comparison of prediction performance} 
\label{tab:perf}
\begin{threeparttable}
\begin{tabular}{lccc}
\toprule
Method & ACC (\%) & FNR (\%) & Inference FLOPs \\
\midrule
FFNN & 21.7 & 4.6 & 1860 \\
DQN & 40.7 & 23.2 & 14520 \\
SARSA-$\lambda$ & 31.1 & 24.7 & 321 \\
IOHMM-BO & \textbf{45.3} & \textbf{5.0} & 297~--~13533\tnote{1} \\
\bottomrule
\end{tabular}
\begin{tablenotes}
\item[1] IOHMM-BO FLOPs range corresponds to model order $K=1$ to $K=6$ (composite state size 2 to 64); for $K=4$ (default), FLOPs are about 2,180.
\end{tablenotes}
\end{threeparttable}
\end{table}
\subsection{Energy Saving Evaluation}
We evaluate net energy consumption using the receiver power model from Section~\ref{subsec:Energymodel}. Table~\ref{tab:energy} shows the normalized energy \(N_{\text{energy}}\) for each method (lower is better). IOHMM-BO achieves a saving of 43\% (\(N_{\text{energy}}=0.57\)), slightly behind DQN (46\% saving, \(N_{\text{energy}}=0.54\)) but significantly better than FFNN (5\% saving) and SARSA-$\lambda$ (19\% saving). However, DQN’s higher saving comes at the cost of a much higher FNR (23.2\%), which may cause frequent missed grants and degrade user experience. IOHMM-BO provides a more balanced trade-off between energy efficiency and reliability.
\begin{table}[htbp]
\centering
\caption{Normalized energy consumption}
\label{tab:energy}
\begin{tabular}{lccc}
\toprule
Method & \(N_{\text{energy}}\) & Saving (\%) \\
\midrule
Reactive DPM & 1.000 & 0\% \\
SARSA-$\lambda$ & 0.81 & 19\% \\
FFNN & 0.95 & 5\% \\
DQN & 0.54 & 46\% \\
IOHMM-BO & \textbf{0.57} & \textbf{43\%} \\
\bottomrule
\end{tabular}
\end{table}
\subsection{Computational Overhead}

Table~\ref{tab:perf} also lists the inference FLOPs per TTI. For $K=4$, IOHMM-BO requires about 2,180 FLOPs, which is higher than FFNN (1,860) and SARSA-$\lambda$ (321) but much lower than DQN (14,520). %On a 300 MHz DSP with 0.5 ms TTI (max 150,000 FLOPs), IOHMM-BO uses only about 1.45\% of the computational budget, comfortably meeting real-time constraints.

\subsection{Sensitivity Analysis: Trade-off between FNR and Energy Saving}

\begin{figure}[htbp]
\centering
\includegraphics[width=0.41\textwidth]{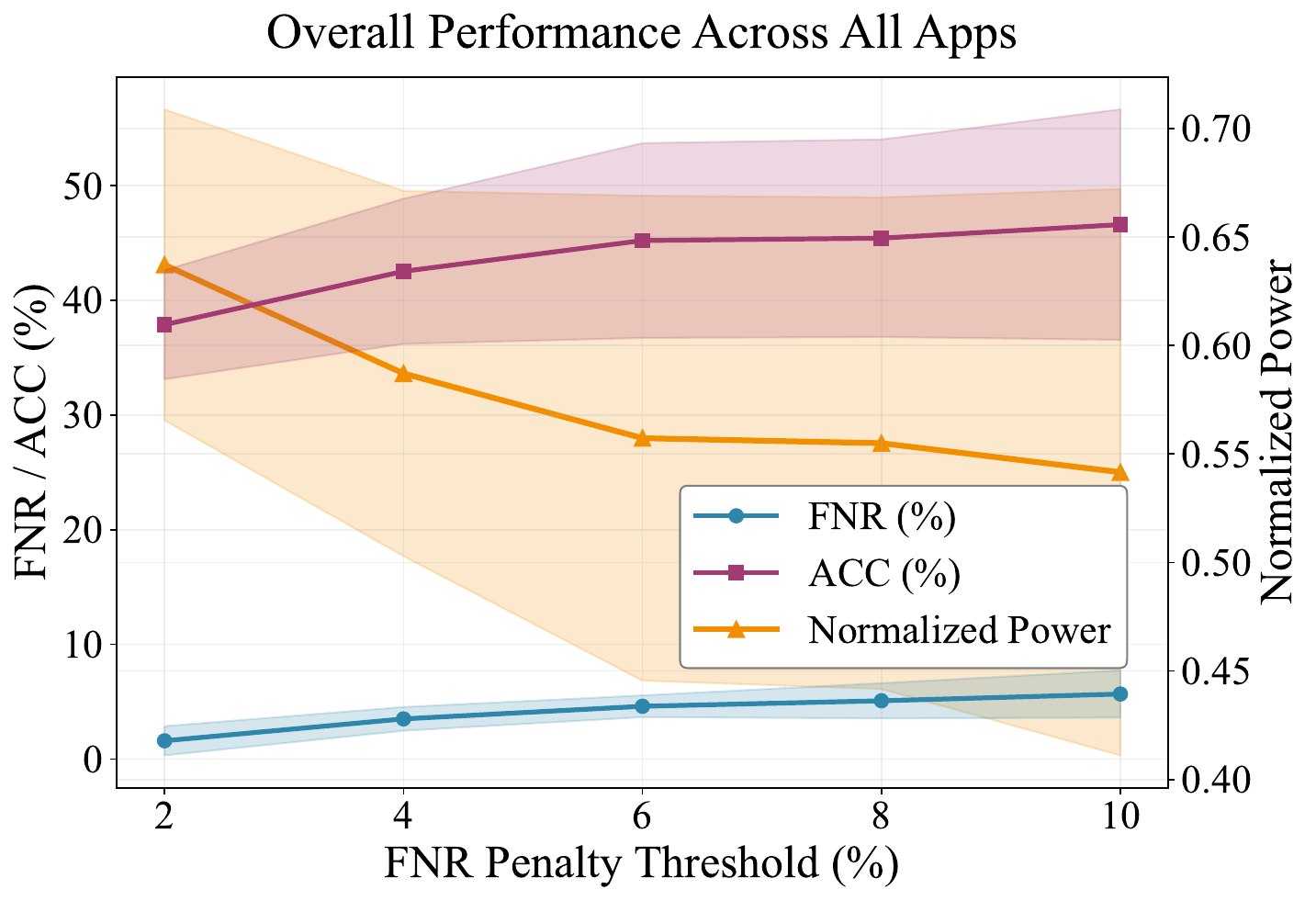}
\caption{Trade-off between false negative rate (FNR) and normalized energy. As the allowed FNR increases, energy consumption decreases, while accuracy slightly improves.}
\label{fig:fnr_energy_tradeoff}
\end{figure}

By varying the penalty threshold \(thr\) in the outer objective, we can explore different operating points of IOHMM-BO. Table~\ref{tab:tradeoff} and Fig.~\ref{fig:fnr_energy_tradeoff} show how average FNR, ACC, and normalized energy change as \(thr\) increases from 0.02 to 0.10.

\begin{table}[htbp]
\centering
\caption{Effect of FNR penalty threshold on performance and energy}
\label{tab:tradeoff}
\begin{tabular}{cccc}
\toprule
FNR threshold & Avg. FNR & Avg. ACC (\%) & Normalized energy \\
\midrule
0.02 & 0.016 & 37.86 & 0.637 \\
0.04 & 0.035 & 42.51 & 0.587 \\
0.06 & 0.046 & 45.20 & 0.557 \\
0.08 & 0.050 & 45.40 & 0.555 \\
0.10 & 0.057 & 46.58 & 0.542 \\
\bottomrule
\end{tabular}
\end{table}

As expected, relaxing the FNR constraint (higher \(thr\)) 
allows the model to adopt more aggressive sleeping policies, reducing 
normalized energy from 0.637 to 0.542 (about 15\% relative improvement). 
Meanwhile, the average FNR increases moderately (from 1.6\% to 5.7\%), and 
accuracy improves slightly (from 37.9\% to 46.6\%), indicating that a too-strict 
FNR penalty may overly constrain the model and hurt overall prediction quality. 
The default setting \(thr=0.08\) gives a 
point with FNR $\approx$ 4.2\%, ACC $\approx$  45\%, and 
\(N_{\text{energy}} \approx 0.57\), consistent with the results in 
Tables~\ref{tab:perf} and~\ref{tab:energy}. This analysis demonstrates 
that the proposed Bayesian optimization framework can effectively tune 
the trade-off between miss-detection risk and energy savings.

\subsection{Summary of Results}

The experimental evaluation demonstrates that:
\begin{itemize}
    \item IOHMM-BO achieves a balanced prediction performance (45.3\% ACC, 5.0\% FNR), outperforming DQN and SARSA-$\lambda$ in terms of reliability, while maintaining competitive energy saving (43\%).
    \item Compared to the best energy-saver DQN (46\% saving), IOHMM-BO offers lower FNR (5.0\% vs. 23.2\%), which is critical for real communication systems where missed grants cause retransmissions and delay.
    \item Compared to other baselines, the computational complexity of IOHMM-BO is relatively low.
    \item Sensitivity analysis shows that the FNR penalty threshold effectively controls the energy-reliability trade-off.
\end{itemize}
These results confirm that the proposed IOHMM-BO method provides an effective and reliable solution for predictive dynamic power management in 5G NR UEs.

\section{Conclusion}

This paper addressed the problem of predictive DPM for 5G NR UE by proposing a high-order input-output hidden Markov model with Bayesian optimization (IOHMM-BO). The main contributions and findings are summarized as follows.

First, based on real 5G NR traces collected from multiple applications and scenarios, we identified two key characteristics of grant sequences: bursty clusters and long-range dependence on scheduling request (SR) events. These observations motivated the use of a high-order IO-HMM to capture the unobservable scheduling state and temporal memory of the grant generation process.

Second, we proposed a complete IOHMM-BO framework. The high-order model was transformed into an equivalent first-order compound state space to enable efficient inference. The model parameters were estimated via the EM algorithm, and the critical hyperparameters—model order \(K\) and monitoring window length \(Wdz\)—were jointly optimized using Bayesian optimization with an accuracy–FNR trade-off objective. An online prediction and adaptive listen
policy was then derived from the trained model.

Third, we conducted extensive experiments on 80 real 5G NR traces. The results showed that IOHMM-BO achieves:
\begin{itemize}
    \item Balanced prediction performance with an accuracy of 45.3\% and a low false negative rate of 5.0\%, outperforming DQN (40.7\% ACC, 23.2\% FNR) and SARSA-$\lambda$ (31.1\% ACC, 24.7\% FNR) in terms of reliability.
    \item Net energy saving of 43\% relative to reactive DPM, slightly behind DQN (46\%) but with much lower FNR, providing a more practical trade-off for real systems.
    \item Low computational overhead compared to other baselines, resulting in lower inherent energy consumption of the prediction algorithm.
\end{itemize}

Sensitivity analysis further demonstrated that the FNR penalty threshold effectively controls the trade-off between miss-detection risk and energy savings: relaxing the threshold from 0.02 to 0.10 reduces normalized energy from 0.637 to 0.542 (about 15\% improvement) while increasing FNR moderately from 1.6\% to 5.7\%.

In future work, we plan to explore several extensions: (i) real-time closed-loop experiments to evaluate the impact of missed grants on higher-layer protocols; (ii) incorporating user behavior and application-specific QoS requirements (e.g., 5QI) to enable differentiated energy-performance trade-offs; (iii) extending the dataset to cover more edge cases (e.g., sudden channel degradation, handover); and (iv) investigating lightweight deep sequential models tailored to extremely imbalanced grant sequences, potentially combined with the statistical structure of IO-HMM.

Overall, IOHMM-BO provides a reliable, energy-efficient predictive DPM for 5G NR UEs, with a better energy–FNR trade-off than DQN and other baselines.

% \section*{Acknowledgment}

% The preferred spelling of the word ``acknowledgment'' in America is without 
% an ``e'' after the ``g''. Avoid the stilted expression ``one of us (R. B. 
% G.) thanks $\ldots$''. Instead, try ``R. B. G. thanks$\ldots$''. Put sponsor 
% acknowledgments in the unnumbered footnote on the first page\cite{IEEEexample:private}.

\bibliographystyle{IEEEtran}
% \bibliography{IEEEabrv,IEEEexample}
\bibliography{IEEEabrv,ref}
\end{document}